\documentclass[11pt]{article}
\usepackage[utf8]{inputenc}
\usepackage[T1]{fontenc}
\usepackage{lmodern}
\usepackage[margin=1in]{geometry}
\usepackage{amsmath,amssymb}
\usepackage{graphicx}
\usepackage{float}
\usepackage{booktabs}
\usepackage{hyperref}
\usepackage{xcolor}
\usepackage{natbib}
\usepackage{enumitem}
\usepackage{placeins}
\usepackage{xurl}
\usepackage{microtype}
\hypersetup{colorlinks=true, linkcolor=blue!70!black, citecolor=blue!70!black, urlcolor=blue!70!black}

\newcommand{\kap}{\kappa}
\newcommand{\kf}{\kappa_f}
\newcommand{\kV}{\kappa_V}
\newcommand{\kstar}{\kappa^{*}}
\newcommand{\gam}{\gamma_n}
\newcommand{\CA}{C_A}

\title{Bounded, Indeterminate, or a Bug:\\ A Condition-Aware Oracle for Differential Testing of SQL Aggregates}
\author{Madhulatha Mandarapu\thanks{madhulatha@samyama.ai} \and Sandeep Kunkunuru\thanks{sandeep@samyama.ai}}
\date{VaidhyaMegha Private Limited, India\\[2pt]\url{https://samyama.ai/}\\[8pt]July 2026}

\begin{document}
\maketitle

\begin{abstract}
Differential database testing compares results across engines and calls a discrepancy a bug. For
floating-point aggregates this is unsound: engines legitimately disagree because floating-point arithmetic is
not associative. Practice patches this with an epsilon; the leading oracles avoid floating point entirely. We
give the oracle this practice lacks, and show its decisive quantity is not the query but
the engine's \emph{algorithm}. Ground truth is the exact rational value of the stored doubles---arithmetic,
not another engine---and each discrepancy is classified exact, bounded, or indeterminate. The relative error
of an aggregate $f$ under an algorithm $A$ obeys
$\mathrm{rel\,err}\le C_A(n,u)\,\kappa_f^{\,p}$, so the
testability boundary---beyond which no oracle can separate a bug from rounding---is
$\kappa^{*}_{f,A}=(1/C_A)^{1/p}$. \textsc{sum} and \textsc{avg} are the linear case $p{=}1$, recovering
$\kappa^{*}=1/\gamma_n$; variance is $p{=}2$ for the one-pass algorithm and $p{=}1$ for Welford. Across eight
engines in four classes the measured exponent recovers each algorithm, and ClickHouse is the \emph{lone}
one-pass engine ($p{=}2.05$); engine-wide, it returns zero standard deviation, \texttt{NaN} correlation and
wrong-sign regression, while every other engine stays exact and the vendor ships the Welford fix. Its variance is untestable at a condition number $10^{6}$ below \textsc{sum}'s, which ordinary
storage conventions (epoch-nanosecond timestamps, tight sensors) cross---there ClickHouse errs by $2100\%$. A
randomised hunt of $360$ tests finds zero anomalies, evidence the oracle is sound. Code and data are public.
\end{abstract}

\section{Introduction}
Differential testing is the workhorse of database correctness research: run the same query on two engines and
treat a difference in results as evidence of a bug. It rests on an oracle---a rule that decides whether two
answers should have agreed. For integers and strings the oracle is equality. For floating-point aggregates
there is no such rule, because floating-point arithmetic is not associative: two engines that combine the
same column in different orders return different doubles, and both are correct implementations of the
aggregate.

The field's response has been to sidestep the question. \citet{coddtest2025} report that constant folding over
floating-point values ``can result in false alarms, which are avoided in practice by eschewing test cases with
small or large floating-point values.'' \citet{squality2024} document that DuckDB's own harness treats two
floats as matching when they differ by less than $1\%$. The oracles of
\citet{rigger2020pqs,rigger2020norec,rigger2020tlp} either restrict aggregates to a single pivot row, target
predicates rather than aggregation, or compare an engine against \emph{itself} by query partitioning, so a
consistent rounding error cancels and is never observed. The same research group authors the modern
differential-testing line, including \citet{sqlxdiff2025}, and it consistently routes around floating point
rather than through it.

An epsilon is not merely a crude oracle; it is wrong in a direction that depends on the data. If the
computation nearly cancels, the true answer is tiny while the intermediate magnitudes are large, and any
implementation may lose most of its significant digits legitimately: an epsilon then reports a bug that is not
there. If it does not cancel, the computation is very accurate, and an epsilon far larger than the achievable
error silently accepts a discrepancy that no rounding can explain. The quantity that separates these regimes
is a \emph{condition number}, and it has not been used in database testing.

\paragraph{The boundary belongs to the algorithm, not the query.} The condition number is necessary but not
sufficient. The relative forward error of an aggregate $f$ computed by an algorithm $A$ takes the form
\begin{equation}
\label{eq:general}
\mathrm{rel\,err} \;\le\; \CA(n,u)\,\kf^{\,p_A},
\end{equation}
where $\kf$ is the condition number of the \emph{function} $f$, $p_A$ is an exponent set by the
\emph{algorithm}, and $\CA$ is its stability constant. The point at which the bound reaches $1$---so that it
admits any discrepancy and no oracle can decide---is
\begin{equation}
\label{eq:kstargen}
\kstar_{f,A} \;=\; (1/\CA)^{1/p_A}.
\end{equation}
For a \emph{linear} aggregate such as \textsc{sum} or \textsc{avg}, $p{=}1$ and $\CA=\gam$, so
$\kstar=1/\gam$: this is the only case examined in prior numeric-testing folklore, and it makes the boundary
look like a fact about summation. It is not. \textsc{variance} computed by the textbook one-pass formula
$\big(\sum x_i^2-(\sum x_i)^2/n\big)/n$ is $p{=}2$; computed by Welford's method it is $p{=}1$
\citep{chan1983,welford1962}. \emph{Same function, same $\kf$, different $(\CA,p)$}---so two engines that run
different variance algorithms have testability boundaries that differ by orders of magnitude, and a
differential test that ignores this will misclassify their disagreement.

\paragraph{Contributions.}
\begin{enumerate}[leftmargin=1.4em,itemsep=1pt]
\item \textbf{A general oracle.} We classify each cross-engine discrepancy as \emph{exact}, \emph{bounded}
  (explained by algorithm $A$), or \emph{indeterminate} (the bound admits any discrepancy). Ground truth is
  the exact rational value of the stored doubles, computed in arbitrary precision; the yardstick is
  arithmetic, not another engine. The decision rule is Eq.~\eqref{eq:general}, instantiated per algorithm
  (\S\ref{sec:frame},~\S\ref{sec:oracle}).
\item \textbf{The testability boundary as an algorithm property}, Eq.~\eqref{eq:kstargen}, a one-line
  consequence of the forward error bound. \textsc{sum}/\textsc{avg} recover $1/\gam$; one-pass variance gives
  $(1/\gam)^{1/2}$, a boundary $\sim\!10^{6}$ times smaller (\S\ref{sec:frame}).
\item \textbf{A measured map, and a taxonomy across engine classes.} The measured error exponent
  \emph{recovers each engine's algorithm}. Across eight engines spanning four classes (OLTP row-store,
  embedded, columnar-OLAP, time-series), ClickHouse is the \emph{lone} one-pass engine ($p{=}2.05$) and every
  other is stable ($p\!\le\!1.06$); the choice is engine-level, spanning its whole moment family (variance,
  covariance, correlation, regression), on which it returns zero standard deviation, \texttt{NaN} correlation
  and wrong-sign regression---while the vendor documents the instability and ships the Welford fix
  (\S\ref{sec:var},~\S\ref{sec:taxonomy}).
\item \textbf{A two-regime result} that inverts the field's stated concern. For \textsc{sum}, real queries sit
  at $\kap\!\approx\!1$ and only fuzzer-generated $\pm$MAX reaches the indeterminate regime. For
  \textsc{variance} the indeterminate regime is reached by \emph{ordinary data under ordinary storage
  conventions}---epoch-nanosecond timestamps, metre-scale coordinates, tight sensors---on which ClickHouse's
  variance is wrong by up to $2100\%$ while the stable engines are exact to eleven digits (\S\ref{sec:real}).
\item \textbf{A soundness check with a stated domain.} A randomised hunt of $360$ in-domain tests finds zero
  anomalies---no engine exceeds its bound where the bound can decide---and locates the oracle's domain of
  validity (normalised, non-overflowing columns) (\S\ref{sec:soundness}).
\end{enumerate}

\paragraph{What we do not claim.} The non-associativity of floating-point aggregation and its repair by a
reproducible datatype are due to \citet{muller2018}; reproducible summation more broadly to \citet{demmel2013};
compensated summation to \citet{kahan1965} and \citet{neumaier1974}. The condition number of the sample
variance and the one-pass-versus-two-pass error split are due to \citet{chan1983} and \citet{welford1962}.
Condition-number-driven oracles for numerical \emph{programs} are due to \citet{explanifloat2025} and its
lineage; we wire that idea to SQL engines, which that work explicitly does not address. Equations
\eqref{eq:general} and \eqref{eq:kstargen} and the boundaries derived from them are elementary consequences of
\citet[Thm.~4.1, \S1.9]{higham2002}; we claim no mathematical depth for them. A high exponent is a documented
\emph{design choice} in an engine, not a bug. Our contribution is the oracle's instantiation for SQL
aggregates, the measurement, and the two-regime result.

\section{The condition number and the testability boundary}
\label{sec:frame}
Let $x_1,\dots,x_n$ be the stored IEEE-754 binary64 values of a column. Every binary64 value is a rational
number, so any aggregate over them has an exact rational (or, for roots, exactly-approximable real) value,
computed in arbitrary precision. Let $u=2^{-53}$ be the unit roundoff and $\gam=nu/(1-nu)$ Higham's factor.

\begin{figure}[H]
\centering
\includegraphics[width=0.82\linewidth]{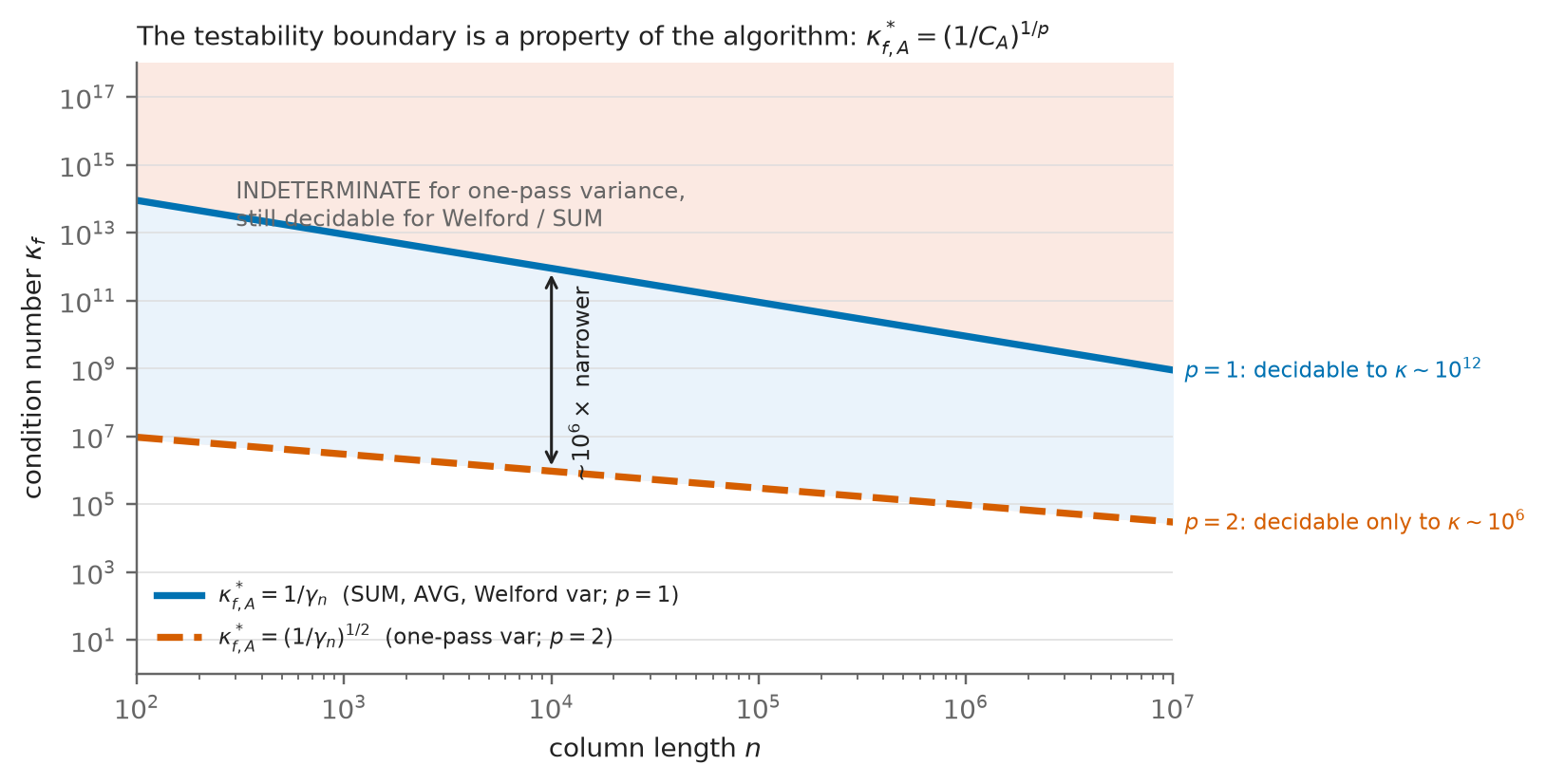}
\caption{Each algorithm sets its own decidable zone: $\kstar_{f,A}=(1/\CA)^{1/p}$. \textsc{sum},
\textsc{avg} and Welford variance ($p{=}1$) stay decidable to $\kap\!\sim\!10^{12}$; textbook one-pass
variance ($p{=}2$) is already indeterminate above $\kap\!\sim\!10^{6}$.}
\label{fig:boundary}
\end{figure}

\paragraph{Summation ($p{=}1$).} For $S=\sum_i x_i$ computed by recursive summation in any order,
$|\hat S-S|\le\gam\sum_i|x_i|$ \citep[Thm.~4.1]{higham2002}. With the condition number of summation
$\kap=\sum_i|x_i|/|\sum_i x_i|$, dividing by $|S|$ gives the relative bound $B=\gam\kap$. Here $\kf=\kap$,
$\CA=\gam$, $p=1$. Note $\kap=1$ whenever the summands share a sign; cancellation is what makes it large.
\textsc{avg} is $S/n$ and inherits the same conditioning (bound $\gam\kap+u$), so it is also $p{=}1$. For
compensated summation \citep{kahan1965,neumaier1974} the bound is $(2u+O((nu)^2))\kap$, whose leading term
carries no $n$.

\paragraph{Variance ($p{=}2$ or $p{=}1$, by algorithm).} For the population variance
$V=\tfrac1n\sum_i(x_i-\bar x)^2$, the relevant condition number is that of the sum of squared deviations
\citep{chan1983,higham2002}:
\begin{equation}
\label{eq:kappaV}
\kV \;=\; \sqrt{\frac{\sum_i x_i^2}{\sum_i (x_i-\bar x)^2}} \;=\; \sqrt{1+\bar x^2/V},
\end{equation}
which is large exactly when the column is \emph{near-constant} (its variance is tiny relative to its raw
second moment). The two algorithms in the wild treat $\kV$ differently:
\begin{itemize}[leftmargin=1.4em,itemsep=1pt]
\item \textbf{Textbook one-pass}, $V=\big(\sum x_i^2-(\sum x_i)^2/n\big)/n$: a single subtraction of two large,
  nearly equal quantities. Its relative error grows as $\CA\,\kV^{2}$ with $\CA\!\sim\!\gam$; the exponent is
  $p{=}2$.
\item \textbf{Welford / two-pass} \citep{welford1962,chan1983}: accumulates deviations from a running mean and
  never forms the cancelling difference. Its relative error grows at most as $\CA\,\kV^{1}$; the exponent is
  $p{=}1$.
\end{itemize}
Both compute the same $V$ from the same data with the same $\kV$; only $(\CA,p)$ differ. By
Eq.~\eqref{eq:kstargen}, their testability boundaries are $\kstar_{\text{Welford}}\approx1/\gam$ and
$\kstar_{\text{one-pass}}\approx(1/\gam)^{1/2}$. Figure~\ref{fig:boundary} plots both: at $n=10^4$ the
one-pass boundary is $9.5\times10^{5}$ against $9.0\times10^{11}$ for the linear case---a decidable window
$\sim\!10^{6}$ times narrower, and one that shrinks only as $\sqrt{nu}$ rather than $nu$.

\section{The oracle: verdicts and two boundaries}
\label{sec:oracle}
Given an engine's answer $\hat y$, the exact value $y$, the condition number $\kf$, and the algorithm $A$ the
engine is measured to use (\S\ref{sec:var}), the oracle emits:
\begin{description}[leftmargin=1.6em,itemsep=1pt]
\item[\textsc{exact}] if $\hat y=y$;
\item[\textsc{indeterminate}] if $B=\CA\,\kf^{\,p}\ge 1$: the bound admits any discrepancy, so \emph{no} oracle
  over this aggregate--algorithm pair can distinguish a bug from rounding;
\item[\textsc{bounded}] if the relative error is at most $B$: the discrepancy is explained by algorithm $A$;
\item[\textsc{anomaly}] otherwise: the discrepancy exceeds what $A$ can produce---a candidate engine bug or a
  modelling error, which must be triaged, never dismissed.
\end{description}
A compensated or Welford variant is judged against its own $(\CA,p)$; judging it by a looser algorithm's bound
would be unsound. Two boundaries follow immediately from $B$.

\paragraph{A1: the testability boundary.} $B\ge 1\iff\kf\ge\kstar_{f,A}=(1/\CA)^{1/p}$
(Eq.~\eqref{eq:kstargen}). Beyond it nothing can be concluded from a discrepancy. Because $\CA$ falls as $n$
grows, \emph{aggregating more rows narrows the regime in which testing can decide anything}; and because $p$
enters as a root, a one-pass algorithm reaches the wall quadratically sooner.

\paragraph{A2: the epsilon crossover.} A fixed relative epsilon $\varepsilon$ misclassifies on both sides of
$\kappa_\varepsilon=(\varepsilon/\CA)^{1/p}$. For $\kf<\kappa_\varepsilon$ we have $\varepsilon>B$ and a real
discrepancy below $\varepsilon$ is accepted (false negative); for $\kf>\kappa_\varepsilon$ we have
$\varepsilon<B$ and legitimate rounding is flagged (false positive). A single fixed epsilon is sound at exactly
one condition number---and for a $p{=}2$ aggregate the safe range on either side is itself quadratically
compressed.

\section{Experimental setup}
\label{sec:setup}
We use five free engines, all local: PostgreSQL~17, MySQL~8.4 and ClickHouse~25.3 in containers, and
DuckDB~1.5.4 \citep{duckdb2019} and SQLite~3.45.1 in-process. Where an engine ships a compensated sum we
measure it (SQLite's \texttt{sum()} is Kahan--Babu\v{s}ka--Neumaier by default, DuckDB offers \texttt{fsum},
ClickHouse offers \texttt{sumKahan}). For variance we call each engine's native population aggregate
(\texttt{var\_pop}, \texttt{varPop}); SQLite ships none, so we register a Welford user-defined aggregate as a
\emph{known-stable reference point} (labelled as such, not a measurement of a SQLite choice). The whole harness
runs on one laptop with no cloud resources, deliberately: a reviewer must be able to re-run it.

Values must reach each engine bit-for-bit or every error measurement is void; we read every row back and
compare bit patterns. Transport is bit-exact in all five engines with one reported exception: SQLite and
PostgreSQL normalise $-0.0$ to $+0.0$ on storage while the others preserve the sign bit; signed zero never
changes a sum or variance. To sweep the summation $\kap$ we generate $n-1$ positive summands with exact sum
$P$ and append one negative value so the total is $t=2P/(\kap+1)$, recomputing $\kap$ exactly. To sweep the
variance $\kV$ we generate an exactly-centred spread and add an offset, so $\bar x/\mathrm{std}$---and hence
$\kV$ by Eq.~\eqref{eq:kappaV}---ranges from $1$ to $10^9$. For real queries we use TPC-H \citep{tpch} at
scale factor $0.1$.

\section{Results I: SUM is the linear baseline}
\label{sec:sum}
\textsc{sum} instantiates the oracle at $p{=}1$ and reproduces the classical picture, which we state briefly as
the baseline the variance results depart from. Across a $(\kap,n,\text{engine},\text{variant})$ grid there were
zero anomalies: no plain sum ever exceeded $\gam\kap$, and every compensated sum stayed under its tighter
bound. At a fixed table size where each engine chooses a sequential plan, DuckDB, PostgreSQL and MySQL return
\emph{bit-identical} plain sums at every $\kap$, so a differential test across those three has no power against
a summation bug; this is a property of the configuration, not the engines, and dissolves at their default
parallel settings. The testability boundary is $\kstar=1/\gam$ ($9.0\times10^{11}$ at $n{=}10^4$), and the
epsilons in use are sound at exactly one $\kap$: a $10^{-9}$ rule crosses at $\kappa_\varepsilon\!\approx\!900$
and DuckDB's $1\%$ at $\approx\!9\times10^{9}$. Real TPC-H sums---including Q9's signed profit---sit at
$\kap\!\approx\!1$, ten orders of magnitude inside the decidable zone, so the shipped epsilons are $15\times$
to $1.5\times10^{8}\times$ too permissive and the practical \textsc{sum} hazard is false negatives. The only
route to \textsc{sum}'s indeterminate regime is fuzzer data: replicating SQLancer's generator, exactly
cancelling $\pm$MAX drives $\kap\!\approx\!10^{306}$, and the fraction of undecidable columns is non-monotone
in $n$ (peaking near $20\%$ at $n{=}1000$). The full \textsc{sum} map is in the artifact; the rest of the paper
concerns what changes when $p\ne 1$.

\section{Results II: variance recovers the algorithm}
\label{sec:var}
\begin{figure}[H]
\centering
\includegraphics[width=0.70\linewidth]{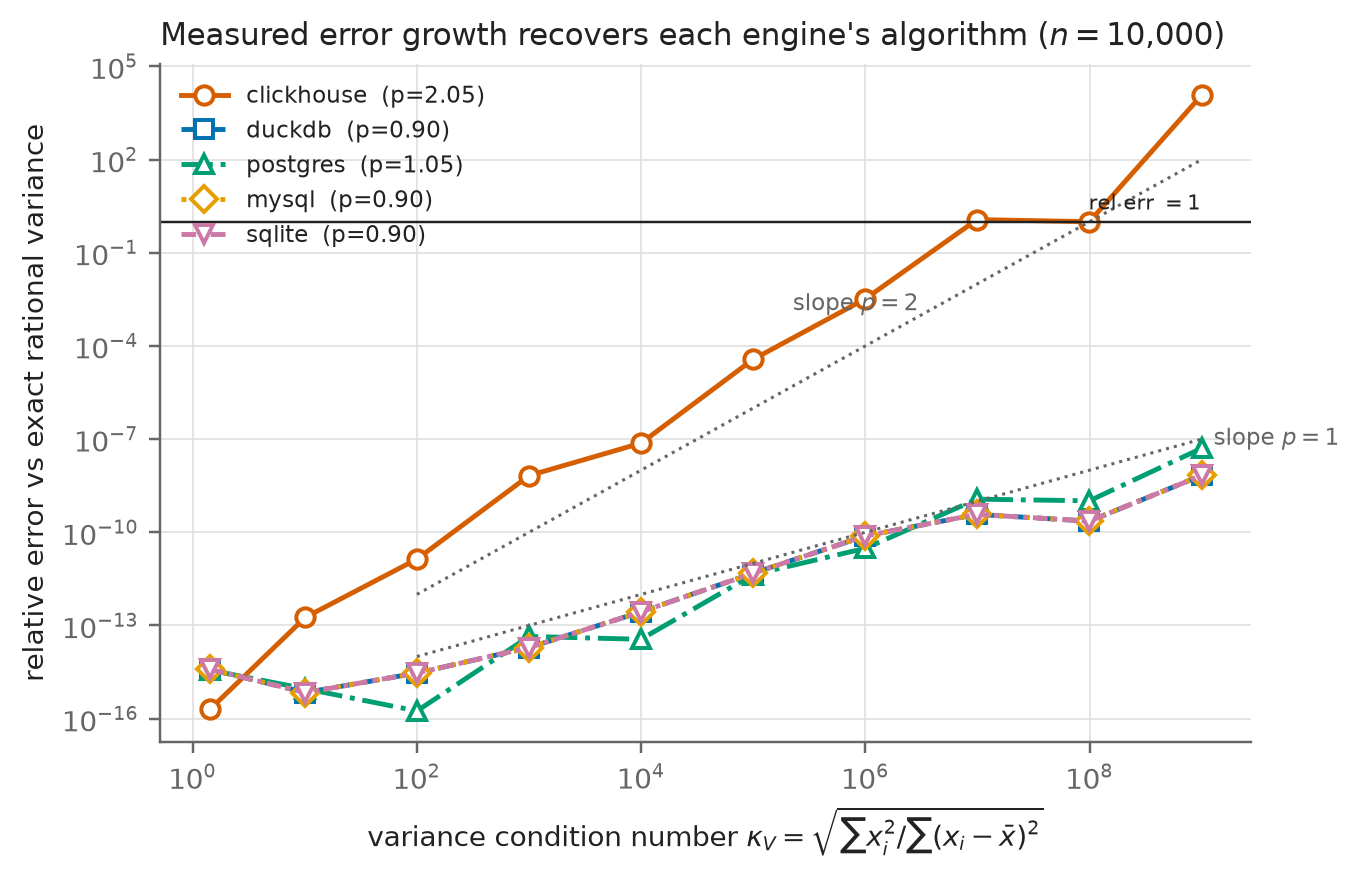}
\caption{Reading the algorithm off the measured slope. ClickHouse ($p{=}2.05$) follows the one-pass slope and
blows past relative error $1$; DuckDB, PostgreSQL, MySQL and the SQLite Welford reference ($p\!\le\!1.05$) stay
far below their linear bound. Ground truth is the exact rational variance.}
\label{fig:exp}
\end{figure}

\paragraph{The measured exponent identifies each engine's algorithm (VE1).} We sweep $\kV$ from $1$ to $10^9$
at $n{=}10^4$ and fit the log--log slope of each engine's relative error against $\kV$. Figure~\ref{fig:exp}
shows a clean split. ClickHouse's \texttt{varPop} tracks slope $p{=}2.05$ and crosses relative error $1$ near
$\kV\!\approx\!10^{7}$: it is the textbook one-pass algorithm, as its exponent alone reveals. The other four
grow at most linearly (measured slopes $0.90$ for DuckDB, MySQL and the SQLite Welford reference, $1.05$ for
PostgreSQL) and never approach their own $p{=}1$ bound, the signature of a stable two-pass or Welford
computation. Recovering the known exponents of the reference Welford implementation ($0.90$) and of one-pass
($2.05$) is also our leak check: a harness that could not read off the algorithm from reference code could not
be trusted on engines.

\begin{figure}[H]
\centering
\includegraphics[width=0.60\linewidth]{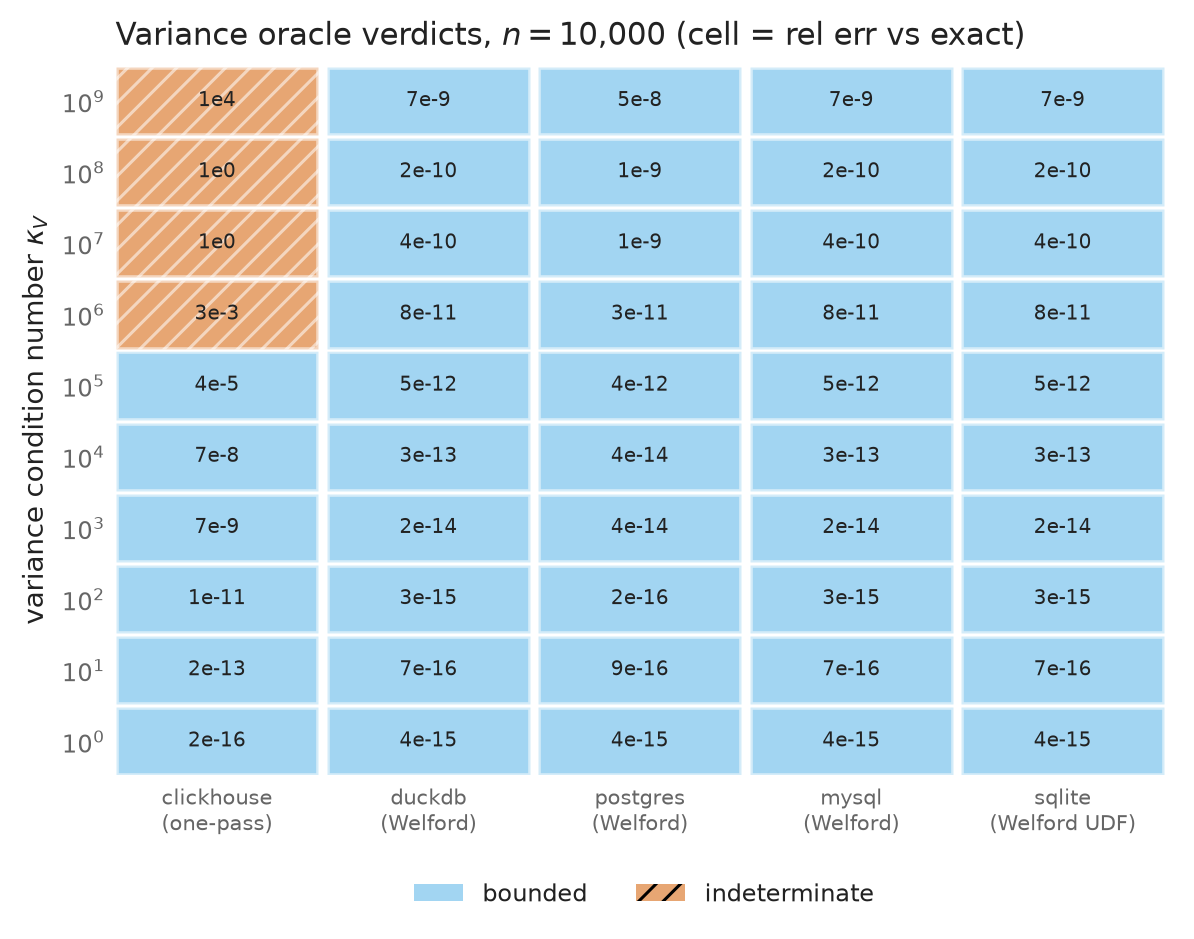}
\caption{Variance oracle verdicts at $n=10{,}000$; each cell is the relative error against the exact rational
variance. ClickHouse (one-pass) is indeterminate for $\kV\ge10^{6}$; the four stable engines remain bounded
through $\kV=10^{9}$. Zero anomalies.}
\label{fig:varmap}
\end{figure}

\paragraph{The classification map (VE2).} Figure~\ref{fig:varmap} shows every verdict. Across all 70 variance
cells (and 70 sum cells) there were \textbf{zero anomalies}: the per-algorithm bound is sound, including the
one-pass bound applied to ClickHouse, which correctly turns \textsc{indeterminate} exactly where
$\gam\kV^{2}\ge1$ rather than falsely accusing the engine. The consequence for differential testing is stark:
above $\kV\!\approx\!10^{6}$, ClickHouse is \textsc{indeterminate} while the four stable engines are still
\textsc{bounded} to eleven digits. A test comparing ClickHouse's variance against DuckDB's on such a column
sees a large disagreement that is entirely rounding on ClickHouse's side---uninterpretable without the $p{=}2$
bound, and misread as a ClickHouse bug by any fixed epsilon tighter than the one-pass error.

\FloatBarrier
\section{Results III: real data versus its representation}
\label{sec:real}
\paragraph{Real analytic columns are safe---for variance too (VE3a).} Table~\ref{tab:ve3a} gives $\kV$ for the
numeric columns of TPC-H, including dates stored as epoch days and seconds. Every one sits at $\kV\in[1.9,13.4]$,
far inside the decidable zone of \emph{both} algorithms. As with \textsc{sum}, raw analytic columns pose no
conditioning hazard, and we do not claim otherwise.

\begin{table}[t]
\centering
\small
\begin{tabular}{lrr}
\toprule
Real TPC-H column (sf$=0.1$) & $n$ & $\kV$ \\
\midrule
\texttt{l\_quantity}                        & 600{,}572 & 2.03 \\
\texttt{l\_extendedprice}                   & 600{,}572 & 1.92 \\
\texttt{l\_extendedprice*(1-l\_discount)}   & 600{,}572 & 1.91 \\
\texttt{l\_shipdate} (epoch days / seconds) & 600{,}572 & 13.4 \\
\texttt{o\_orderdate} (epoch days)          & 150{,}000 & 13.3 \\
\bottomrule
\end{tabular}
\caption{Real analytic columns are well-conditioned for variance ($\kstar_{\text{one-pass}}=9.5\times10^{5}$
at these $n$). Multiplicative unit changes leave $\kV$ invariant.}
\label{tab:ve3a}
\end{table}

\begin{figure}[H]
\centering
\includegraphics[width=0.86\linewidth]{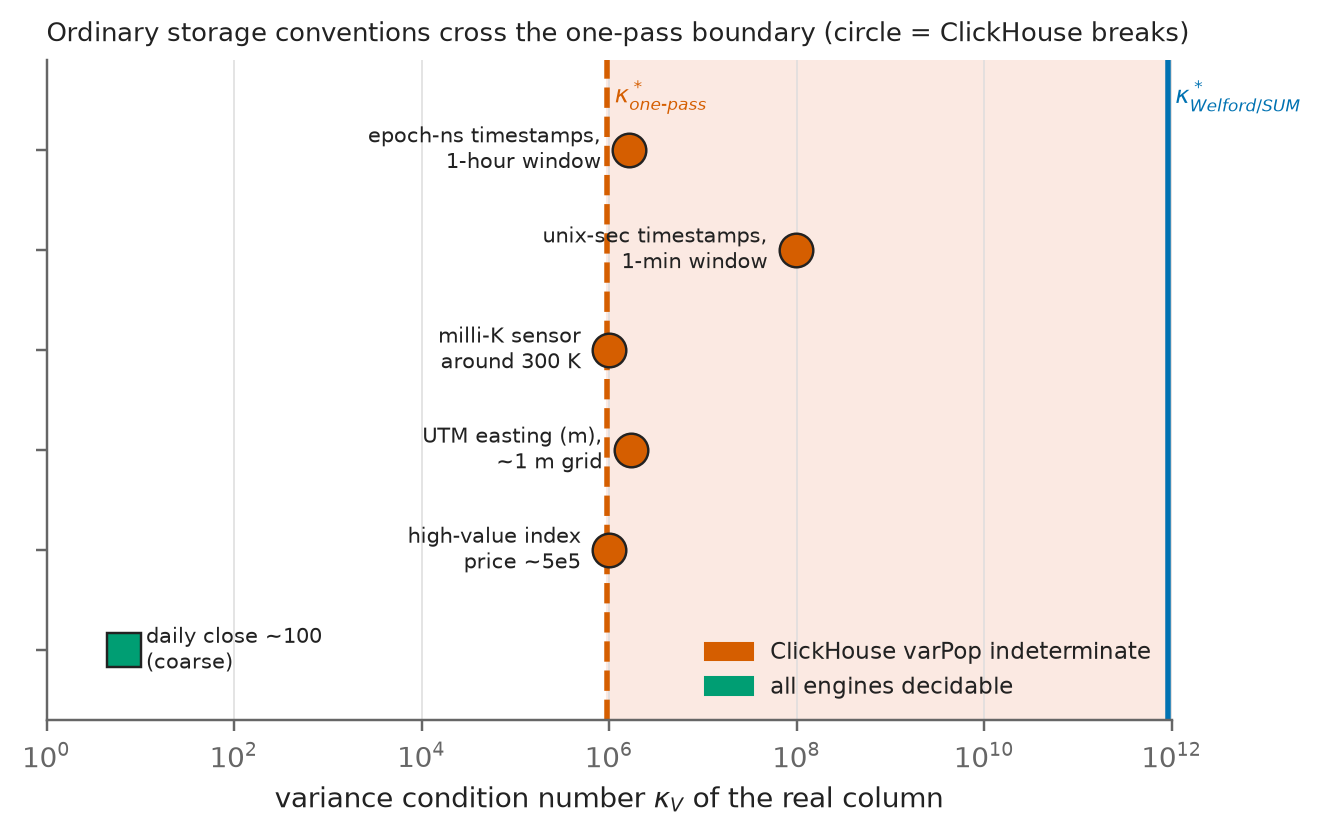}
\caption{Six real storage conventions against the two boundaries. Five cross $\kstar_{\text{one-pass}}$ (shaded)
and break ClickHouse's variance; the coarse control does not. All six are far below
$\kstar_{\text{Welford/SUM}}$.}
\label{fig:real}
\end{figure}

\paragraph{But variance conditioning is offset-sensitive, and ordinary schemas cross the boundary (VE3b).}
Unlike the summation $\kap$, which is invariant to a shift of the data, $\kV=\sqrt{1+\bar x^2/V}$ grows with
the additive \emph{offset} a schema chooses. Real systems routinely store a fine-grained quantity as an
absolute value against a distant zero: timestamps as epoch nanoseconds (the default of pandas
\texttt{datetime64[ns]}), temperatures in kelvin, positions as projected metre coordinates, prices of
high-value near-constant instruments. For such a column, $\bar x/\mathrm{std}$---and hence $\kV$---is enormous
even though the underlying spread is perfectly ordinary. Table~\ref{tab:ve3b} takes six named conventions,
builds a real float64 column for each, computes $\kV$ exactly, and runs it end-to-end through all five engines.
Five of the six cross $\kstar_{\text{one-pass}}=9.5\times10^{5}$; on all five, ClickHouse's \texttt{varPop} is
indeterminate or grossly wrong---up to a relative error of $21$ (a $2100\%$ error) on one-minute event
timestamps stored as unix seconds---while DuckDB, PostgreSQL, MySQL and the SQLite reference remain accurate to
$10^{-10}$. The sixth, a coarse daily price series ($\kV=6.8$), stays decidable everywhere: the honest negative
control. Figure~\ref{fig:real} places all six against the two boundaries.

\begin{table}[t]
\centering
\footnotesize
\begin{tabular}{lrccc}
\toprule
Real storage convention ($n=10{,}000$) & $\kV$ & ClickHouse & DuckDB & PG/MySQL/SQLite \\
\midrule
unix-sec timestamps, 1-min event window     & $9.8\times10^{7}$ & \textsc{indet.} ($2.1\times10^{1}$) & $4\times10^{-10}$ & bounded \\
UTM easting (m), $\sim$1\,m survey grid      & $1.7\times10^{6}$ & \textsc{indet.} ($2\times10^{-2}$)  & $1\times10^{-10}$ & bounded \\
epoch-ns timestamps, 1-hour window          & $1.6\times10^{6}$ & \textsc{indet.} ($1\times10^{-2}$)  & $1\times10^{-10}$ & bounded \\
milli-kelvin sensor around 300\,K           & $1.0\times10^{6}$ & \textsc{indet.} ($8\times10^{-3}$)  & $3\times10^{-11}$ & bounded \\
high-value index price $\sim\!5\times10^{5}$ & $9.9\times10^{5}$ & \textsc{indet.} ($5\times10^{-3}$)  & $3\times10^{-11}$ & bounded \\
\emph{daily close $\sim$100 (coarse control)} & $6.8$            & bounded                             & bounded           & bounded \\
\bottomrule
\end{tabular}
\caption{Ordinary storage conventions push real columns across the one-pass testability boundary. ClickHouse's
one-pass \texttt{varPop} becomes indeterminate (parenthesised relative error against the exact variance) while
the stable engines stay accurate. Verdicts are \textsc{indeterminate}, never \textsc{anomaly}: the $p{=}2$
bound admits the error, so it is not a false accusation of ClickHouse.}
\label{tab:ve3b}
\end{table}

This inverts the field's concern in a sharper way than \textsc{sum} did. For \textsc{sum} the indeterminate
regime was reached only by fuzzer-generated $\pm$MAX; here it is reached by a log table of event timestamps.
The exposure is not exotic data but an everyday representation choice interacting with an engine's algorithm,
and it is $10^{6}$ times more likely under a one-pass engine.

\FloatBarrier
\section{Results IV: the taxonomy across engine classes and the moment family}
\label{sec:taxonomy}
Two questions decide how far the variance result generalises: is one-pass a widespread choice or one engine's,
and does it stop at variance? We answer both by measurement.

\begin{figure}[H]
\centering
\includegraphics[width=0.86\linewidth]{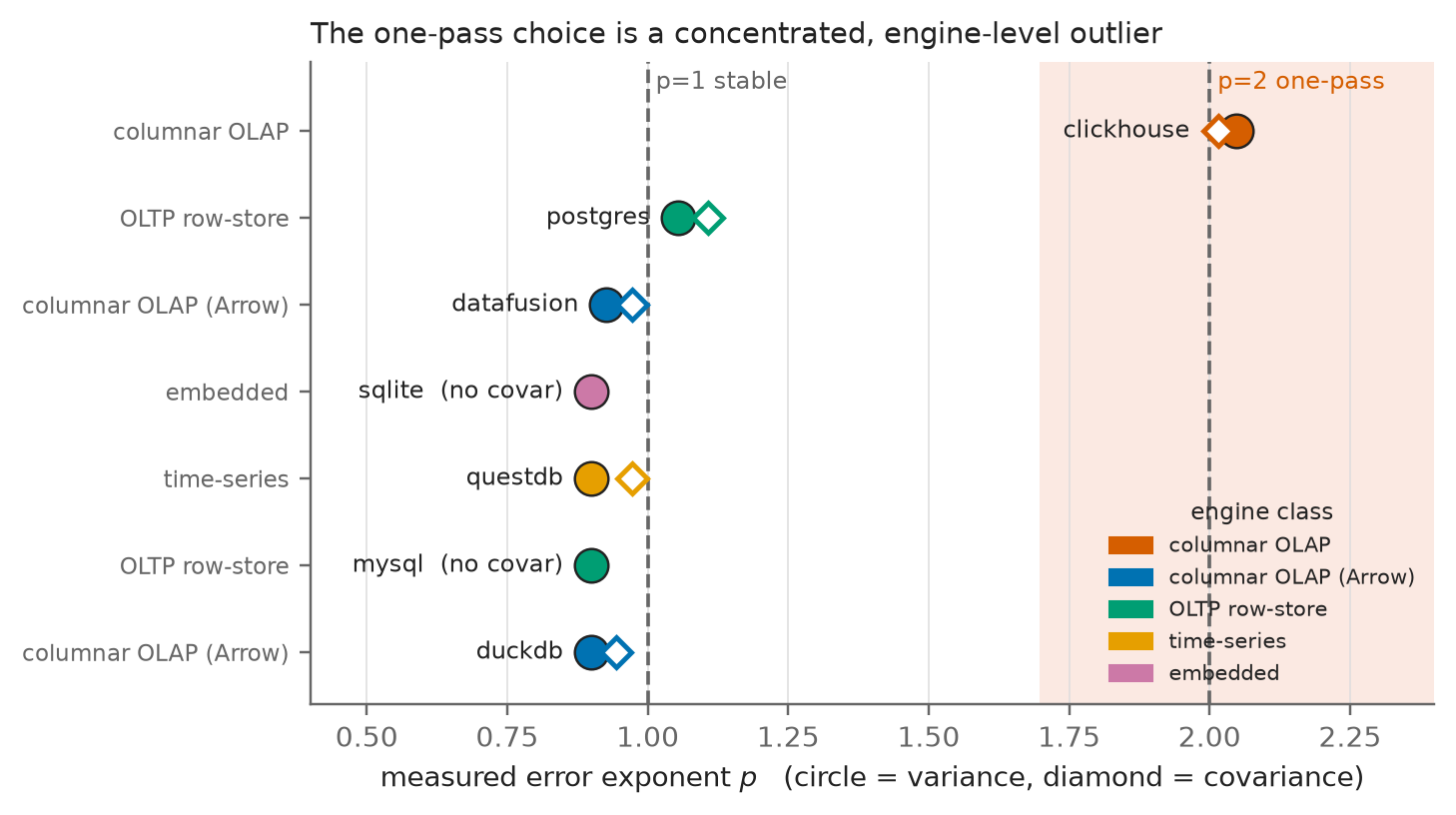}
\caption{Engine-class taxonomy of the measured exponent, for variance (circles) and covariance (diamonds)
across eight engines and four classes. ClickHouse alone sits in the one-pass region; every other engine, in
every class, is stable. The one-pass choice is a concentrated, engine-level outlier.}
\label{fig:taxonomy}
\end{figure}

\paragraph{One-pass is a concentrated outlier across four engine classes.} We widen the survey with three
engines: DataFusion (Arrow columnar), QuestDB (time-series, whose core columns are near-constant timestamps),
and MonetDB (classic columnar). Fitting the variance exponent on all of them (Figure~\ref{fig:taxonomy}) leaves
ClickHouse the \emph{only} one-pass engine ($p=2.05$); PostgreSQL, MySQL, DuckDB, SQLite, DataFusion and QuestDB
are all stable ($p\le 1.06$). The time-series engine is stable, so the timestamp hazard of \S\ref{sec:real} is a
property of ClickHouse's algorithm, not of time-series systems. MonetDB is excluded: its client's
\texttt{binary64} read-back is not bit-exact ($1388$ of $4000$ values differ by one ulp), so its transport fails
our exactness check and its cells are void---reported, not hidden.

\paragraph{The one-pass choice spans the whole moment family.} We repeat the measurement for the SQL covariance
family---\texttt{covar\_pop}, \texttt{corr}, \texttt{regr\_slope}, \texttt{stddev} (PostgreSQL, DuckDB,
ClickHouse, DataFusion and QuestDB expose it; MySQL and SQLite have no native covariance). ClickHouse's
\texttt{covarPop} fits $p=2.02$, matching its variance; every other engine is again stable. So the one-pass
algorithm is an \emph{engine-level} choice, not a per-aggregate one. Its consequences on a near-constant column
($\kappa_V\approx10^{8}$, well past the one-pass boundary) are not merely large errors but qualitatively broken
answers (Table~\ref{tab:family}): ClickHouse returns a covariance wrong by $6200\%$, a correlation of
\texttt{NaN}, a regression slope of the \emph{wrong sign} ($-3.6$ where the true slope is $+0.60$), and a
standard deviation of exactly $0$---reporting a varying column as constant---while the stable engines stay
accurate to eleven digits.

\begin{table}[H]
\centering
\small
\begin{tabular}{lrrr}
\toprule
Aggregate ($\kappa_V\approx10^{8}$, $n=4000$) & exact & ClickHouse (one-pass) & stable engines \\
\midrule
\texttt{covar\_pop}  & $0.588$  & $36.9$ ($6200\%$ error)        & ${\sim}10^{-8}$ \\
\texttt{corr}        & $0.517$  & \texttt{NaN}                   & ${\sim}10^{-9}$ \\
\texttt{regr\_slope} & $+0.599$ & $-3.6$ (\emph{wrong sign})     & ${\sim}10^{-8}$ \\
\texttt{stddev\_pop} & $0.991$  & $0$ (\emph{reports constant})  & ${\sim}10^{-9}$ \\
\bottomrule
\end{tabular}
\caption{ClickHouse's uniformly one-pass moment family produces qualitatively broken results on an ordinary
large-offset column; every stable engine is accurate. These verdicts are \textsc{indeterminate} (past
$\kstar_{\text{one-pass}}$), not \textsc{anomaly}.}
\label{tab:family}
\end{table}

\paragraph{The vendor documents the instability and ships the fix.} This is not a hidden defect: ClickHouse
documents \texttt{varPop}/\texttt{covarPop}/\texttt{stddevPop} as numerically unstable and provides
\texttt{varPopStable}, \texttt{covarPopStable} and \texttt{stddevPopStable} as slower, lower-error alternatives
\citep{clickhousedocs}. We confirm the \texttt{*Stable} variants are the Welford fix the oracle prescribes: on
the same $\kappa_V\approx10^{8}$ column, \texttt{varPopStable} and \texttt{covarPopStable} give relative errors
$1.5\times10^{-8}$ and $8.6\times10^{-9}$, versus $100\%$ and $6200\%$ for the defaults. The oracle's role is to
give the \emph{boundary} at which the default becomes untestable and the stable variant becomes necessary.

\FloatBarrier
\section{Soundness: a bug hunt and the domain of validity}
\label{sec:soundness}
An oracle earns trust by never accusing a correct engine. We stress it with a randomised hunt: seven generators
(well-conditioned, cancelling, mixed-magnitude, integer, near-constant, subnormal and near-overflow) over three
table sizes and four seeds, run through six engines for \textsc{sum}, \textsc{avg}, variance and ClickHouse's
\texttt{varPopStable}, classifying every result against its algorithm's bound. An \textsc{anomaly} is an error
\emph{exceeding} the bound at a decidable condition number---a candidate bug. Over $360$ in-domain engine-tests
there were \textbf{zero anomalies}: no engine, default or stable, exceeded its bound where the bound can decide.
This is evidence that the bounds are sound, not that the engines are bug-free.

The hunt also located the oracle's \emph{domain of validity}, which we state rather than paper over. Higham's
bound assumes normalised arithmetic and no overflow; two generators violate those assumptions and are excluded.
Under gradual \textbf{underflow} (subnormal inputs) the deviations $x_i-\bar x$ round to zero and every engine
returns variance $0$: the normalised-float bound does not hold, and every engine agrees, so this is a modelling
limit, not a bug. Under \textbf{overflow} (values near $10^{300}$) the exact $\sum x_i^2$ exceeds the
\texttt{binary64} range and the result is genuinely $\pm\infty$, so the bound is undefined. The oracle applies to
columns that are normalised and whose exact intermediate sums are finite---which is the entire real-data regime
of \S\ref{sec:real}.

\FloatBarrier
\section{Related work}
\label{sec:related}
\paragraph{Differential and metamorphic DBMS testing.} \citet{rigger2020pqs} verify a single pivot row, which
restricts multi-row aggregates; \citet{rigger2020norec} rewrites predicates; \citet{rigger2020tlp} partitions a
query against the same engine, so a consistent rounding error cancels rather than being detected.
\citet{coddtest2025} avoid floating-point test cases explicitly; \citet{squality2024} record the $1\%$ epsilon
used in practice; \citet{sqlxdiff2025} test emerging engines against relational references with an exact-result
oracle. None classifies a numeric discrepancy against a forward error bound, none uses a condition number, and
none observes that the testability boundary depends on the engine's aggregation algorithm.

\paragraph{Numerical analysis and reproducibility.} The summation bound and the framing of $\kap$ are standard
\citep{higham2002,goldberg1991}; compensated summation is due to \citet{kahan1965,neumaier1974} and reproducible
summation to \citet{demmel2013}. The condition number of the sample variance and the one-pass-versus-two-pass
error analysis are due to \citet{chan1983} and \citet{welford1962}. \citet{muller2018} solve non-determinism
\emph{within} an engine with a reproducible aggregation datatype---complementary to our problem of deciding
whether two \emph{different} engines should have agreed. \citet{explanifloat2025} build a condition-number-aware
oracle for standalone numerical programs and state that it does not address database systems. Our contribution
is to carry the general form $\CA\kf^{\,p}$ into DBMS differential testing, instantiate it for a $p{=}1$ and a
$p{=}2$ family, and measure where real engines and workloads land.

\section{Limitations and honest negatives}
\label{sec:limits}
\begin{itemize}[leftmargin=1.4em,itemsep=1pt]
\item \textbf{Real analytic columns are safe.} We measured this (Table~\ref{tab:ve3a}) and report it. The
  variance hazard arises from a storage \emph{representation}, not from analytic data as such.
\item \textbf{A high exponent is a design choice, not a bug.} ClickHouse's one-pass moment family is a
  documented speed/accuracy trade-off, and the engine ships \texttt{*Stable} variants (\S\ref{sec:taxonomy}).
  Our claim is only that a differential test must judge the default against the $p{=}2$ bound; every verdict
  against it here is \textsc{indeterminate}, never \textsc{anomaly}.
\item \textbf{We found no engine bug.} A randomised hunt of $360$ in-domain tests (\S\ref{sec:soundness})
  produced zero anomalies: evidence the oracle is sound, not that the engines are correct. A targeted
  bug-finding campaign across engine versions remains future work.
\item \textbf{The exponent is measured, not proven per engine.} We read $p$ from a log--log fit over a
  controlled sweep; vendor source corroborates the ClickHouse and Welford assignments but the reported finding
  is the measurement. The stable engines' sub-linear fitted slopes ($0.90$--$1.05$) mean only that they never
  approach their $p{=}1$ bound, not that variance is unconditionally safe on them.
\item \textbf{Differential power depends on configuration.} As for \textsc{sum}, bit-identity and parallel
  nondeterminism are configuration-relative; any ``engine X cannot detect engine Y'' claim is qualified by
  thread and plan settings.
\item \textbf{Scope.} We measure \textsc{sum}, population \textsc{variance}/\textsc{stddev}, and the covariance
  family (\texttt{covar\_pop}, \texttt{corr}, \texttt{regr\_slope}) over binary64 (\S\ref{sec:taxonomy});
  \textsc{avg} is not measured separately but follows analytically as $S/n$, inheriting \textsc{sum}'s $p{=}1$
  conditioning up to one rounding. Sample variance differs only by the $n/(n-1)$ factor and shares the
  algorithm's exponent. The bound's domain excludes subnormal and overflowing columns (\S\ref{sec:soundness}),
  and \textsc{decimal} semantics, windowed aggregates and \textsc{group by} partitioning are untouched.
\end{itemize}

\section{Conclusion}
Floating-point aggregates have been a blind spot for differential database testing, patched with an epsilon
whose soundness was never examined. Treating the exact rational value as ground truth and a forward error bound
as the decision rule turns the epsilon into a computable object and yields a testability boundary
$\kstar_{f,A}=(1/\CA)^{1/p}$ that belongs to the engine's \emph{algorithm}, not merely the query. \textsc{sum}
is the benign linear case; variance is not, and the one-pass algorithm that one popular engine ships makes its
variance untestable a million times sooner than its sum---on data as ordinary as a table of timestamps. The
community has been guarding the door its fuzzers open while leaving open the one its schemas walk through.

\section*{Artifact availability}
Code, the exact workloads, and a one-command reproduction are available at
\url{https://github.com/samyama-ai/numeric-semantics-oracle}. All experiments run on a single machine with free
software and no cloud resources.

\bibliographystyle{plainnat}
\bibliography{paper21_numeric_oracle}
\end{document}